%% file: main.tex
\documentclass[pdflatex,sn-mathphys-num]{sn-jnl}
\usepackage[T1]{fontenc}
\usepackage{hyperref}
\usepackage{paralist}
\usepackage{sistyle} % for numbers 
\usepackage{tcolorbox}
\usepackage{xurl}
\usepackage{graphicx}%
\usepackage{multirow}%
\usepackage{amsmath,amssymb,amsfonts}%
\usepackage{amsthm}%
\usepackage{mathrsfs}%
\usepackage[title]{appendix}%
\usepackage{xcolor}%
\usepackage{textcomp}%
\usepackage{manyfoot}%
\usepackage{booktabs}%
\usepackage{algorithm}%
\usepackage{algorithmicx}%
\usepackage{algpseudocode}%
\usepackage{listings}%
\usepackage{pifont}% http://ctan.org/pkg/pifont
\newcommand{\cmark}{\ding{51}}%
\usepackage{cleveref}
\newcommand{\quotes}[1]{``#1''} %for qoutation marks
 
\newcommand{\autorefappendix}[1]{\hyperref[#1]{Appendix~\ref*{#1}}}  % to change subsubsection 
\usepackage[edges]{forest}
\usetikzlibrary{arrows.meta,shadows.blur}
\usepackage{pdflscape}
\usepackage{tcolorbox}
\usepackage{eso-pic}
\theoremstyle{thmstyleone}%
\theoremstyle{thmstyletwo}%

\theoremstyle{thmstylethree}%

\begin{document}
\AddToShipoutPictureFG*{%
\put(0,760){%
\makebox[\paperwidth]{%
\begin{tcolorbox}[
colback=gray!10,
colframe=black,
boxrule=0.3pt,
arc=1mm,
width=0.85\textwidth,
center,
boxsep=1mm
]
This document is the author's pre-print for a paper published at the \textit{Information and Computer Security}. Please, refer to the journal's final open-access version:  https://doi.org/10.1108/ICS-12-2025-0541
\end{tcolorbox}
}}
}

\title{Users' Activity Logs: the Good, the Bad, the Misconception, and the Disastrous}
\author{\fnm{Eman} \sur{Alashwali}}\email{ealashwali@kau.edu.sa}
\affil{\orgname{King Abdulaziz University}, \orgaddress{\city{Jeddah}, \country{Saudi Arabia}}}

%\markboth
%{Author \headeretal: Preparation of Papers for IEEE TRANSACTIONS and JOURNALS}
%{Author \headeretal: Preparation of Papers for IEEE TRANSACTIONS and JOURNALS}

\abstract{Most service providers, such as Google, save logs from data generated by users while using the service. Many service providers provide users with privacy controls to manage whether, how, and for how long the data is saved and used by the service provider. While most prior studies focused on the negative side of users’ activity logs, such as users' lack of awareness about the logs' privacy controls and users' privacy concerns toward their data, this work aims to provide a balanced view of users’ perceptions regarding activity logs by considering the positive, negative, and extremely negative (hence disastrous) sides, as well as the misconceptions of activity logs. In this work, we present a case study of Google's Activity controls by conducting a secondary analysis of interview data from 30 Google personal account holders in Saudi Arabia. Using template analysis, we analyzed the data from the lens of four main themes: the good, the bad, the misconception, and the disastrous aspects of users’ activity logs from the users’ perspective. Our findings uncover new themes and use cases, offering a balanced view of users’ perceptions of activity logs, and provide a better understanding and a useful source for subsequent studies on related topics. We conclude with practical recommendations for service providers, privacy researchers and experts, and users alike.}
\keywords{security, privacy, policy, settings, data, Google, web, applications}

\maketitle

%If you do not have or do not want to include a photo, you can use IEEEbiographynophoto as shown below:
\input{sec/intro}

\input{sec/related}

\input{sec/method}

\input{sec/results}
\input{sec/discussion}
\input{sec/conclusion}
\input{sec/ack}

\bibliography{ref}
\clearpage
\input{sec/appendices}
\end{document}

%% file: sec/intro.tex
\section{Introduction}
Most web and mobile service providers collect and analyze data generated by users while they use their services, such as users' search queries, visited locations, and watched videos within a given service. In this paper, we refer to such data as \quotes{\textbf{activity logs}}. Activity logs contain data that varies in sensitivity by application and data type. For example, intuitively, activity logs of orders in a food delivery application are unlikely to contain surprising or sensitive information to the user. However, with the rapid advancements in the Internet of Things (IoT), smart home devices, behavioral analytics, and personalization, many service providers are now saving unconventional types of activity logs, e.g., saving every search query the user typed in a search engine, or every voice command given to a smart speaker, over a long period of time. 

Many service providers offer users privacy controls (also known as privacy settings) that allow them to determine whether, how, and for how long their activity logs are saved and used by the service provider. However, these controls are often hidden~\cite{habib20,chen19}, confusing even for security and privacy experts~\cite{germain22,green18}, and have relatively permissive defaults~\cite{chen19,haselton17}. Users who are unaware of default settings and how to adjust them may inadvertently share data~\cite{liu11,madejski12,shih15,green18}, leading to consequences such as embarrassment, bullying, stalking, identity theft, and fraud~\cite{gross05}. Some service providers have been accused of misleading users through privacy setting interfaces. For example, in the US in 2022, Google agreed to a \$392 million settlement for misleading users about privacy controls that failed to clearly inform them of the status of the location-tracking setting~\cite{kang22}. 

Previous work has investigated users' perceptions and behaviors with respect to privacy controls in various contexts, such as social media~\cite{gross05,ellison07,liu11,madejski12,habib22}, smart home devices~\cite{malkin19,zheng18,lau18}, mobile apps~\cite{ramokapane19,felt12}, and web service providers such as Google~\cite{alashwali24,farke21,haque23}. While most prior studies focused on the negative side of activity logs, such as users' lack of awareness about privacy controls and users' privacy concerns about their data, this work aims to provide a broader perspective of users' perceptions regarding activity logs. 

\textit{Our research aims to answer the following question: what are the positive and negative aspects, and misconceptions about activity logs from the users' perspective?} 

To answer this question, we conducted a case study on Google's \textit{Activity controls}. We qualitatively analyzed interview data from 30 Google personal account holders in Saudi Arabia regarding their views on Google's \textit{Activity controls}, which provide users with control over the activity logs that Google saves about them while using Google services. 

Our work provides a secondary analysis of data obtained from a previous study by Alashwali and Cranor~\cite{alashwali24}. Nevertheless, the analysis we provide in this work was not covered in~\cite{alashwali24}, and is an original contribution for this paper. Namely, in~\cite{alashwali24}, the focus of the analysis was on identifying themes related to users' awareness, use, preferences, behaviors, and concerns about Google's \textit{Activity controls} and the data it collects about them, as well as the steps they take to control Google's collection or use of this data. This paper complements Alashwali and Cranor's work~\cite{alashwali24} by analyzing data from a new perspective, focusing on the positive and negative aspects and misconceptions regarding Google's Activity controls as a case study of activity logs, from the users' perspective. See the related work in~\autoref{sec:perception} for further details about how this work is different from~\cite{alashwali24}.

Our results show that participants identified some positive aspects of Google's \textit{Activity controls}, such as saving browsing history and bookmarks, synchronization, security and safety, and parental control. On the other hand, they also identified negative aspects, such as a lack of knowledge of Google's data practices and the presence of Google's \textit{Activity controls}, privacy concerns, and biased web content. Some Google activity logs' features, such as tailored ads, were controversial, where some participants appreciated them while others did not. We also identified misconceptions among participants, such as confusion between local browsing history and cloud-based activity logs, and between security and privacy measures. Participants also raised serious concerns about Google's \textit{Activity controls}, which can lead to serious consequences, such as unauthorized logins and data breaches.

The main contribution of this paper is offering a balanced view of users’ perceptions of activity logs and providing a better understanding and a useful source for subsequent studies on related topics. We offer a set of practical recommendations for service providers, security and privacy researchers and experts, and users alike.

%% file: sec/related.tex
\section{Background and Related Work} \label{sec:related}
In this section, we first describe Google's \textit{Activity controls}. We then summarize related work on privacy perceptions and behaviors associated with privacy controls. Finally, we provide a background on the Saudi society and privacy. 

Before we move to the related work, we need to shed light on the difference between two main types of privacy controls that have been described in the literature~\cite{alashwali25}: between a user and other users (\quotes{\textbf{user-to-user}}) and between a user and institution(s) (\quotes{\textbf{user-to-institution}}). Our study falls into the second category. 

% Google's activity controls
\subsection{Google's \textit{Activity controls}} \label{sec:activity_logs} 
Google's \textit{Activity controls}\footnote{Now called \textit{Personalisation and activity}.}~\cite{google21}, an area inside Google accounts that spans from the \textit{Data \& personalization}\footnote{Now called \textit{Data and privacy.}} menu, was introduced in 2016 and then called \textit{My Activity}~\cite{giles16,farke21}. It provides Google account holders with control over multiple activity logs while using Google services, namely: the \textit{Web \& App Activity}, \textit{Location History}\footnote{Now called \textit{Timeline}.}, and \textit{YouTube History}, with the following descriptions provided by Google (retrieved in 2021~\cite{google21}, see~\autoref{fig:options} for illustration): 
%--------------- BEGIN FIGURE -----------------
\begin{figure}
	\begin{center}
	\includegraphics[width=\columnwidth]{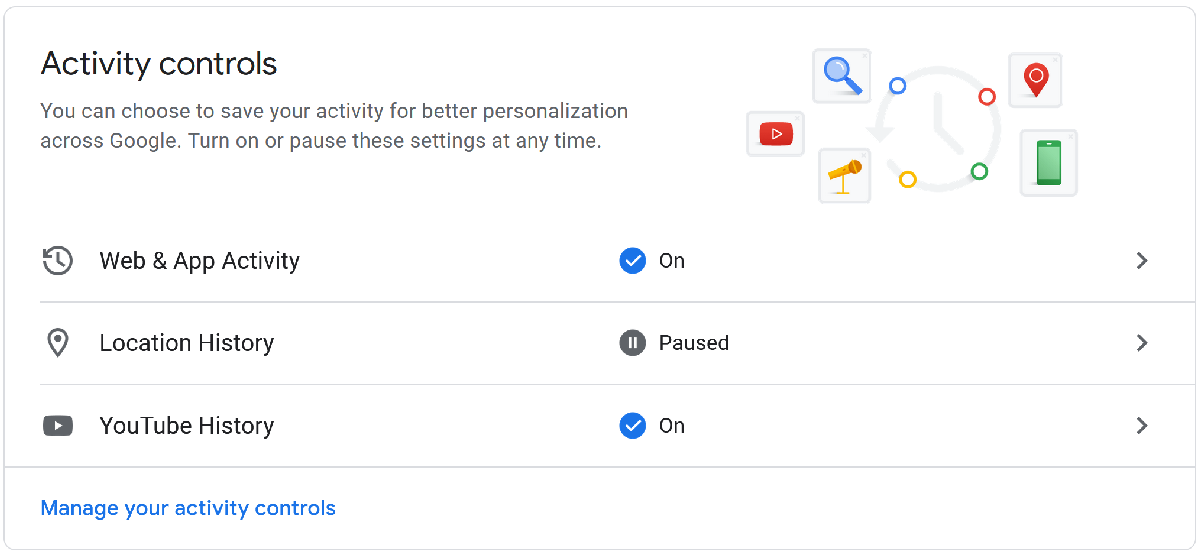}
    \end{center}
	\caption{The Google's \textit{Activity controls} lists three types of activity that Google saves: \textit{Web \& App Activity} (enabled by default), \textit{Location History} (disabled by default), and \textit{YouTube History} (enabled by default)~\cite{google21}.}
	\label{fig:options}
	%\vspace{-4mm}
 \end{figure}
%--------------- END FIGURE ----------------- 
\begin{enumerate}
	\item \textit{Web \& App Activity}: \quotes{Saves your activity on Google sites and apps, including associated info like location ...}
	\item \textit{Location History}: \quotes{Saves where you go with your devices, even when you aren't using a specific Google service ...}
	\item \textit{YouTube History}: \quotes{Saves the YouTube videos you watch and the things you search for on YouTube ...}
\end{enumerate}

As of 2021~\cite{google21,alashwali24}, when a user creates a Google account, the \textit{Web \& App Activity} and the \textit{YouTube History} are enabled by default. In contrast, the \textit{Location History} is disabled by default. The default retention periods for the \textit{Web \& App Activity}, \textit{Location History}, and \textit{YouTube History} are 18 months, 18 months, and 36 months, respectively. Users can adjust the retention period by changing the \textit{Auto-delete} setting, which can be set to decrease or increase up to \quotes{Don't auto-delete activity}. Users can view and manually delete some or all of their activity logs. They can also customize which data to save in the logs. For example, in the \textit{Web \& App Activity}, users have two check boxes: \quotes{Include Chrome history and activity from sites, apps, and devices that use Google services}, which is checked by default, and \quotes{Include audio recordings}, which is unchecked by default. Similarly, in the \textit{YouTube History}, there are two choices: \quotes{Include the YouTube videos you watch} and \quotes{Include your searches on YouTube}, which are both checked by default. 

We chose to base our study on interview data from Google's \textit{Activity controls} because Google is one of the largest service providers in the world. In 2024, Google had 1.8 billion active Gmail users~\cite{kiran24,kumar24}. 

\subsection{Perceptions and Behaviors of Privacy Controls} \label{sec:perception}
Earlier work on the perceptions and behaviors of privacy controls has focused on user-to-user controls, such as profile and post visibility settings on social media platforms. While awareness and use of user-to-user privacy controls on social media platforms were low in the early years of social media use~\cite{gross05}, subsequent studies on social media privacy have shown that awareness and use of user-to-user privacy controls have significantly improved over time~\cite{stutzman12}. However, several researchers have warned that overlooking user-to-institution privacy controls may lead to an illusory sense of control over privacy~\cite{brandimarte13,heyman14}. That is, when users' perceived control is limited to a specific aspect of their data, ignoring what Stutzman et al. called \quotes{silent listeners}, such as service providers and third parties~\cite{haque23,brandimarte13,stutzman13}. This highlighted the need for further research to understand users' perceptions and behaviors for user-to-institution privacy controls, which our work tries to contribute to. 

In recent years, several studies have looked at the user-to-institution privacy controls from the users' perspective, such as~\cite{alashwali24}, which we extend and reuse its data for the secondary analysis of this work. While we use the same data, this work differs from~\cite{alashwali24} in the analysis template, and hence in the emerged themes presented in the results and discussion sections. In~\cite{alashwali24}, Alashwali and Cranor conducted interviews with 30 Google personal account holders in Saudi Arabia about perceptions and behaviors of Google's \textit{Activity controls}. Their study was exploratory in nature, with an overarching goal to bridge the knowledge gap on privacy perceptions and behaviors in non-Western societies, such as the Saudi society, and to compare patterns with similar studies conducted in the US. Their analysis was focused on the following main themes: awareness, use, preferences, behaviors, and concerns about Google's \textit{Activity controls}, where they reported the identified sub-themes at a fine-grained level. In the awareness theme, the authors reported the awareness level (how many participants reported they were aware of Google's \textit{Activity controls}) and identified sub-themes for how they became aware of them. Similarly, in the use theme, they reported use level (how many participants reported they used Google's \textit{Activity controls}) and identified sub-themes for the reasons for using or not using them. In the attitudes part, the authors reported participants' feelings when they viewed some of their data in their accounts (e.g., \textit{Web \& App Activity}) in an interactive task during the interview, and why they felt that way. The authors also recorded participants' attitudes towards Google's default retention period and their preferences for the ideal retention period, and their attitudes and preferences towards the acceptable use of their data by Google. Finally, the authors reported participants' privacy concerns. Additionally, they discussed the observed trends on privacy awareness, attitudes, preferences, concerns, and behaviors of Saudi participants compared to similar studies on US participants. 

Our work extends Alashwali and Cranor's work~\cite{alashwali24} by looking at the data from a new lens focused on identifying and reporting: what are the positive (good), negative (bad), and extremely negative (disastrous) aspects, and the misconceptions regarding activity logs from the user's perspective, using data on Google's \textit{Activity controls} as a case study.  

Most of the studies in user-to-institution privacy controls showed a lack of, or vague, awareness and use of these types of controls. For example, Farke et al.'s study on the then-called Google's \textit{My Activity} found that only 12\% of participants said they were \quotes{extremely aware} of Google's \textit{My Activity}, while the majority were either \quotes{moderately aware} or \quotes{somewhat aware} of it~\cite{farke21}. They also found that only 36\% of participants visited Google's \textit{My Activity} page~\cite{farke21}. Nevertheless, the study showed that when participants were exposed to Google's \textit{My Activity}, they were more likely to report feeling less concerned about data collection and to perceive it as more beneficial~\cite{farke21}. Malkin et al's. study on Google and Amazon smart speakers found that less than half of smart speaker users knew that their audio recordings were permanently saved in their smart speakers, while 41.4\% incorrectly believed that their recordings were saved temporarily~\cite{malkin19}. Additionally, they found that only a minority used the smart speaker's privacy controls~\cite{malkin19}. For example, of those who knew about the logs' deletion feature, 67.9\% never used it~\cite{malkin19}. Lau et al.'s study on smart speaker users found that most users did not use their devices' privacy controls~\cite{lau18}. In Ul Haque et al.'s study, which was based on Google's \textit{Activity controls}, the majority of participants kept the default settings~\cite{haque23}.

Some of the documented reasons for the lack of awareness and use of user-to-institution privacy controls include: lack of knowledge about the controls, discoverability issues (hidden controls), overtrusting or undertrusting the service provider, and resignation~\cite{habib20,chen19,ramokapane19,zheng18,lau18}.

Research showed that participants were surprised when they viewed their activity logs and saw the data saved about them. Farke et al. reported that more than a third of participants were surprised by the amount of data collected about them~\cite{farke21}. Malkin et al. reported that many participants were surprised that their voice interactions with the smart speaker were permanently saved~\cite{malkin19}. Similarly, multiple studies on mobile apps reported that most participants were surprised by the amount~\cite{jung12} and frequency~\cite{balebako13} of data collected by their mobile apps, as well as the destinations~\cite{balebako13} to which it was sent. While most service providers' default settings are configured to retain activity logs for long periods, research shows that most users prefer shorter retention periods than the defaults~\cite{malkin19}. 

Research showed that privacy concerns are highly contextual. Studies by Lau et al.~\cite{lau18}, Zheng et al.~\cite{zheng18}, and Zeng et al.~\cite{zeng17} on smart home devices reported that participants had few privacy concerns. Reasons for low concern include not feeling targeted, trusting the manufacturer, an incomplete understanding of the privacy risks associated with the service, and a trade-off between privacy and convenience or functionality~\cite{lau18,zheng18,zeng17}. Malkin et al. reported that only 28.3\% of participants had privacy concerns about their smart speakers. Still, they also noted that users raised greater concerns in specific contexts, such as when audio recordings contain children's or guests' voices or if recordings are used for ads~\cite{malkin19}, offering contextual integrity as an explanation~\cite{nissenbaum04,malkin19}.

% Saudi Users
\subsection{Saudi Society and Privacy} \label{sec:saudi}
Saudi Arabia is a developing country in southwestern Asia, established in 1932~\cite{wikipedia25}. It has one of the youngest populations in the world. As of 2022, 39\% of the population was below 30 years old~\cite{gas22_pop}. It has around 17\% of the world’s proven oil reserves and is the second-largest member of the Organization of the Petroleum Exporting Countries (OPEC)~\cite{opec25}. This situation contributed to rapid development and technology adoption. The Internet penetration rate in Saudi Arabia reached 99\% in 2023~\cite{cst23}. As of 2022, it is reported that over 80\% of the Saudi population are social media users~\cite{kemp22,blogger22}. 

As of 2021, when Alashwali and Cranor's~\cite{alashwali24} interview study was conducted (we reused its data for this study), Internet users in Saudi Arabia have no legal protection for their personal data. However, the Personal Data Protection Law (PDPL) was enforced in Saudi Arabia in September 2024~\cite{malin24,yates24}.

%% file: sec/method.tex
\section{Method}
In this section, we describe our methods. We first describe the interview data we used in our study. Then, we describe how we analyzed the data. Finally, we list some limitations.

\subsection{Interviews} \label{sec:interview}
We obtained 30-participant interview data from Alashwali and Cranor's study on privacy perceptions and behaviors of Google personal account holders in Saudi Arabia~\cite{alashwali24}. Participants were recruited through the first author of~\cite{alashwali24} via her university's mailing list, social and professional WhatsApp groups, and social media accounts, including Twitter and LinkedIn. Full details of the recruitment process are provided in~\cite{alashwali24}. The interviews were conducted online in August 2021 in Arabic (the native language of the participants and the interviewer) in a semi-structured format. Each interview lasted 33 minutes on average and included open-ended, multiple-choice, and interactive task-based questions. In the latter, the interviewer asked participants to log in to their Google accounts on their computers and to answer questions about their settings, awareness of Google's \textit{Activity controls}, and their sentiments after viewing their activity logs. The interview questions are provided in~\autorefappendix{app:questions}.

\subsection{Participants}
All participants were living in Saudi Arabia. Participants included 19 (63\%) females and 11 (37\%) males, with ages ranging from 18 to 54 years. Eight participants (27\%) reported having technical backgrounds (i.e., a university degree or work in Computer Science (CS), Information Systems (IS), Information Technology (IT), or Computer Engineering (CE)). None of them reported a degree or work in the cybersecurity field (confirmed via a question at the end of the interview). See~\autoref{tab:demo} in~\autorefappendix{app:demo} for further demographic details.

\subsection{Ethical considerations}
The secondary use of Alashwali and Cranor's interview data~\cite{alashwali24} was approved by the Institutional Review Board (IRB) committee of the main author of~\cite{alashwali24}, King Abdulaziz University. The data was obtained directly from the authors of~\cite{alashwali24}. All participants were adults who aged 18 years or older and provided consent before taking part in the study. Participants were informed about the purpose of collecting the data and that the results will be published in anonymized format. 

\subsection{Analysis} \label{sec:analysis}
We qualitatively analyzed the transcribed interview data using template analysis, a style of thematic analysis that combines inductive and deductive approaches, with an emphasis on hierarchical coding but without a specific prescription for the number of levels required or what the levels represent~\cite{king25,brooks14}. It involves creating a codebook (the template), which defines the themes identified by the researcher in the dataset, and can be updated during the analysis where codes can be added, updated, and deleted. The researcher normally starts the analysis with a \quotes{a priori themes}, which represent themes anticipated by the researcher.

We started the analysis with the following a priori themes: the good, the bad, the misconception, and the disastrous aspects of Google's activity logs, which were inspired by previous work~\cite{alashwali24}. Then, for each interview transcript, the researcher first read the entire transcript and identified the emergent sub-themes of the a priori themes and documented the frequency and supporting narrative(s) of each sub-theme. In this paper, we translated the interview questions and participants' direct quotes from Arabic into English, using the Google Translate~\cite{translate26} followed by a manual review by the researcher, who is a native Arabic speaker and fluent in English. 

While multiple coders can confirm consistent interpretations, analyzing the data by a single experienced researcher is deemed acceptable in both the template analysis method~\cite{king25} and Human-Computer Interaction (HCI) research~\cite{mcdonald19}.

%% file: sec/results.tex
\section{Results} \label{sec:results}
\begin{figure}[!t]
\centering
\resizebox{0.95\textwidth}{!}{%
\forestset{
	direction switch/.style={
		forked edges,
		for tree={
			align=left,
			edge+=thick, 
			font=\sffamily,
		},
		where level=1{minimum width=13em}{},
		where level<=1{draw=black}{},
		where level>=1{folder, grow'=0}{},
	},
}
	\newlength\gap
	\setlength\gap{10mm}

\begin{forest} 
%============= level 0 ================
		direction switch
		[Google's Activity controls 
%============= level 11 ================
			[The Good
				[History \& Bookmark]
                [Synchronization]
                [Tailored Content]
                [Parental Control]
                [Security and Safety]
                    %[Reasons for preferred product\\(\autoref{tab:preferred_reason})]
				%]
			] % end prefs.
%============= level 12 ================ 
			[The Bad
                [Lack of Knowledge]
                [Privacy Concerns]
                [Bias]
                [Ads]
			] % end attitude
%============= level 13 ================
			[The Misconception
				[Local vs{.} Online Activity Logs]
                [Security vs{.} Privacy Measures]
                [Meaning of Defaults]
			] %end awareness
%============= end================
%============= level 14 ================
			[The Disastrous
                [Technology Abuse]
                [Unauthorized Login]
                [Data Breach]
			] % end actionability
		]
	\end{forest}
}
\caption{A diagram summarizing our qualitative analysis themes for the positive (good), negative (bad), and extremely negative (disastrous) aspects, and misconceptions of Google's \textit{Activity controls} from the users' perspective.}
\label{fig:diagram}
\end{figure}
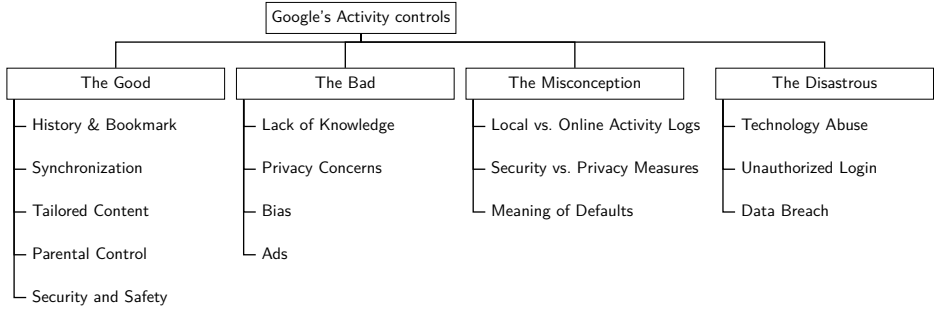

In this section, we first summarize the participants' demographics. We then summarize the four overarching themes and subthemes that categorize users' perceptions of Google's \textit{Activity controls}, including \textit{Web \& App Activity}, \textit{YouTube History}, and \textit{Location History}. Namely, we summarize the positive (good), negative (bad), and extremely negative (disastrous) aspects, as well as misconceptions, of Google's activity logs.~\autoref{fig:diagram} summarized the subthemes we identified under the main themes.

%=========== BEGIN FIGURE ======================
%=========== BEGIN FIGURE ======================

We denote participants as (P\_\#). Since our study is qualitative in nature and aims to identify themes rather than quantify them, we do not report the number of participants reporting each theme. Instead, we adopt the following terminology from Emami-Naeini et al.~\cite{emami19}: (few: 0\% -- 25\%; some: 25\% -- 45\%; about half: 45\% -- 55\%; most: 55\% -- 75\%; almost all: 75\% -- 100\%). 

\subsection{Positive Aspects of Google's \textit{Activity controls} (the Good)} \label{sec:good}
In this section, we summarize the subthemes we identified as positive aspects of Google's \textit{Activity controls} (the good).

% good aspects of activity logs
\subsubsection{History and Bookmark} 
Some participants found the \textit{Web \& App Activity} useful for retrieving previously visited websites or mistakenly closed websites, as described by P\_04: \quotes{For example, there is something I want to remember, so I can see it again and review it}. To some participants, the \textit{YouTube History} was especially useful as P\_11 noted: \quotes{On YouTube, if it is up to me, I would never turn on the auto-delete}. A few participants appreciated the \textit{Location History} feature for saving location maps of infrequently visited locations that they might otherwise forget. 

\subsubsection{Synchronization}
One participant mentioned that she finds the \textit{Web \& App Activity} useful to synchronize the web history between multiple devices: \quotes{Because I use them from several devices, for example, sometimes I search on Google Chrome from my mobile phone, sometimes from the desktop computer, sometimes from the laptop, so here it is like a unified place for all activities from all devices} (P\_08).

\subsubsection{Tailored Content} \label{sec:tailored_content}
Some participants perceived the tailored content that Google's \textit{Activity controls} may provide positively. This was mentioned in relation to ads, YouTube videos, or web browsing content. The main perceived benefits were the convenience, relevance to their interests, and ease and speed of finding web and ad content. P\_03 can summarize this: \quotes{It might make it easier for me and save time for them to find me important things that suit me} and P\_11: \quotes{they help me by giving me things I've searched for before, things that suit me personally}. 

Some participants provided more specific examples of how tailored content helps them, such as P\_08, a postgraduate student who said: \quotes{I am studying for a doctorate ...  so I search a lot, so the results come out to me faster and better, and it understands more, so I reach the things I actually want}. Similarly, P\_19, a university instructor, said: \quotes{as a teacher, I prepare [for classes] ... I write the thing, right away it remembers things that I have done before, and things come up for me}. P\_05 mentioned that tailored ads helped her \quotes{find the products on a cheaper site}. A few participants noted that tailored ads are more convenient, such as P\_08: \quotes{because if I leave it not personalized for me, it shows me a lot of annoying or inappropriate advertisements}. This is especially true for conservative societies, such as Saudi society, which may have reservations about the content they view online.

\subsubsection{Parental Control}
A few participants mentioned the use of Google's \textit{Activity controls} for parental control, as P\_09 mentioned using \textit{YouTube History} for this purpose: \quotes{I put the tablet on the same account, the mobile phone on the same account, and the laptop on the same account, to see the videos my children watch}. P\_30 found the \textit{Web \& App Activity} useful for a similar purpose: \quotes{I can go back to the history so that I can see what, for example, my children watched}.  

\subsubsection{Security and Safety}
One participant (P\_08), with a technical background, mentioned that she reviews logs from time to time for security purposes and described: \quotes{to make sure that no one else or another device is open with my account}. P\_19 found a safety benefit related to the \textit{Location History}, namely in location tracking: \quotes{to the contrary, so that you don't get lost, even let someone know where you are}.

%====================================================
\subsection{Negative Aspects of Google's \textit{Activity controls} (the Bad)} \label{sec:bad}
In this section, we summarize the subthemes we identified as negative aspects of Google's \textit{Activity controls} (the bad).

\subsubsection{Lack of Knowledge}
Most participants showed a lack of, or incomplete, knowledge about Google's data practices, including the consent they gave to Google and the existence of Google's \textit{Activity controls} and their features, such as reviewing, deleting, and auto-deleting activity logs. This is exemplified in P\_06's response when he was asked whether he was aware of Google's \textit{Activity controls}: \quotes{I expected something like this to exist, but I don't know it}. He was asked about his feelings when he saw his \textit{Web \& App Activity} data: \quotes{I was surprised that it was stored. I mean, I didn't expect ... that it would record like this}. P\_09 commented on the consent: \quotes{Agree and click click ... but you don't know what the mechanism is, because they give you the terms and conditions on 30 pages}.
 
\subsubsection{Privacy Concerns}
Some participants expressed privacy concerns after they learned about Google's \textit{Activity controls} or after they viewed their activity logs, such as the \textit{Web \& App Activity} and the \textit{YouTube History}, mainly because they felt being watched or tracked, as P\_03 described: \quotes{The feeling of being monitored is honestly unpleasant. That it might record what I do, what I search for, where I am, even YouTube, even if it's not important, I don't like things to be recorded}. P\_03 also mentioned that she might be self-aware of what she does on the web: \quotes{I feel like I might be more careful about what I search for because it's all recorded, I mean it's all saved and recorded in the history, so it's a bit scary}. P\_10 explained concerns about personalized web content in remote work settings: \quotes{in our remote work, we share the screen now ... If you are searching for something, it becomes visible, it immediately shows up to the users, meaning that it shows up to the employees with you. Sometimes it is something personal, meaning that it is not something that you share with people}. 

Regarding retention periods, when asked about Google's defaults, most participants did not prefer Google's default retention periods, which are: 18 months for the \textit{Web \& App Activity}, 18 months for the \textit{Location History}, and 36 months for the \textit{YouTube History}. 

Moreover, a few participants mentioned concerns about data sales and data abuse. When participants were asked about the acceptable use of their \textit{Web \& App Activity} data in three scenarios, most found that using their data to display ads, whether by Google or by third parties with Google partnerships, is unacceptable. 

\subsubsection{Bias}
One participant (P\_10) mentioned that tailored content provides her with a non-neutral (biased) view of the web, such as showing her content based on her search interests, as she described: \quotes{sometimes there is information hiding. It may not be related to what I searched for, but may be useful to me ... meaning that I did not see the subject in its complete form}. Another participant (P\_15) explained how repeated tailored ads can drive people into buying something they do not need: \quotes{you are curious about a particular product, and you don't need it. Then this [Google] comes and makes you an ad, and you feel like you want to buy it because you see it so much}.

\subsubsection{Ads} 
Some participants viewed ads in a generally negative tone, describing them as \quotes{annoying}. P\_14 added: \quotes{if I want something, I will go and search for it myself.} While tailored ads were perceived as a positive feature by some participants (see~\autoref{sec:tailored_content}), others perceived tailored ads negatively, especially when they were repeated as P\_10 said: \quotes{I searched for a specific topic, everything starts showing up related to [it]} or when ads are from third-party websites as P\_14 described: \quotes{you [Google] are taking information from me everywhere, you [Google] are not leaving me alone}. P\_07 criticized the quality of ads and found them useless: \quotes{In 10 years, I haven't gotten an ad that really deserves me to go in and see it}. P\_14 viewed ads as a plan to push users toward premium subscriptions: \quotes{flood YouTube with ads and sell a premium subscription. I mean the plan is simple, but effective}. 

%====================================================
\subsection{Misconceptions about Google's \textit{Activity controls} (the Misconception)} \label{sec:misconception}
In this section, we summarize the subthemes we identified for misconceptions about Google's \textit{Activity controls} (the misconception).

\subsubsection{Local versus Online Activity Logs}
Multiple participants reported confusion between activity logs saved on a local device (e.g., the local browser's \textit{History}) and those saved in the cloud (e.g., Google's \textit{Web \& App Activity}). For example, when asked about the steps they took to protect their privacy on the Internet while using Google services, if any, participants mentioned: deleting the browser's cookies and local history, using private browsing (a feature in most mainstream browsers that deletes local browsing data at the end of the session, e.g., Mozilla's Firefox private browsing~\cite{mozilla25} and Google's Chrome Incognito mode~\cite{incognito25}), and using a Virtual Private Network (VPN). However, while they partially protect privacy, these techniques cannot protect the user's privacy if the service provider saves and uses the user activity logs on the cloud, such as Google's \textit{Web \& App Activity}, \textit{YouTube History}, and \textit{Location History}.

\subsubsection{Security versus Privacy Measures}
While security and privacy measures are interrelated and often conflated, they are technically distinct~\cite {bambauer13}. Generally speaking, privacy concerns the control of who can access and use users' data, whereas security concerns the implementation of technical mechanisms that mediate access or use requests to the data~\cite{bambauer13}. That is, \quotes{security implements privacy}~\cite{bambauer13}. A few participants showed confusion between security and privacy. For example, when asked about the steps they took to protect their privacy on the Internet while using Google services, if any, participants reported security-related measures, such as not using public Wi-Fi (P\_08), changing their password (P\_17), and using two-factor authentication (P\_25).  

\subsubsection{Meaning of Defaults}
 Most participants in this study did not change default settings, which are often permissive by default. For example, Google's \textit{Web \& App Activity} and \textit{YouTube History} were enabled by default. When asked about the reasons for keeping the defaults, some participants indicated misconceptions. For example, not changing defaults unless something went wrong, as P\_13 noted \quotes{I don't go to the account and its choices except when I face a problem, and I search for the problem I want and that's it}. P\_06 said he did not change the defaults because \quotes{It won't have an impact or make a difference in use}, while P\_17 indicated fear of messing things up: \quotes{I don't want to change anything, leave it as it is as long as the email is working. I don't want to change anything and mess with the email. I'm afraid that [I] do something and then something happens to [my] email}. These responses indicate misconceptions about what the default settings are and the effect of changing them.

%====================================================
\subsection{Eextremely Negative Aspects of Google's \textit{Activity controls} (the Disastrous)} \label{sec:disastrous}
In this section, we summarize the subthemes we identified for extremely negative aspects of Google's \textit{Activity controls} (the disastrous).

\subsubsection{Technology Abuse}
While none of the participants reported using Google's \textit{Activity controls} for stalking, we inferred potential abuse from two use cases mentioned by the participants. The first was parental control, where two participants found the \textit{Web \& App Activity} useful for monitoring what their children viewed online. The second was related to location tracking for security and safety, where one participant found the \textit{Location History} useful if someone knew her location in case she got lost. While these sound like good uses of Google's \textit{Activity controls} if they are used by a legitimate party who has been authorized by the account owner or who has the authority to access the account, such as parents of minors, these features can be abused if the account falls into the wrong hands, e.g., for stalking a close person. For example, technology abuse by intimate partners has been documented and studied~\cite{freed18}. Google's \textit{Activity controls}, which allow the account holder's online and location activity logs to be saved, is a feature that can be abused.

\subsubsection{Unauthorized Login} %accoutn hijacking
Participants recognized that Google's \textit{Activity controls} include highly personal data, including \textit{Web \& App Activity}, \textit{YouTube History}, and \textit{Location History}. A few participants mentioned that unauthorized access to their Google account (account hijacking) represents a major privacy concern, as P\_19 expressed: \quotes{I'm afraid that someone might be able to access my account and see what I'm searching for}. P\_19 added that she is more concerned about \quotes{Someone I know more than Google}. A hijacked account can endanger not only their privacy, but even their physical security, e.g., through the \textit{Location History} as P\_23 expressed concerns \quotes{if ... my email was hacked, for example, and the person who accessed my email and activity control would be able to see all the locations that I was there, right?}

\subsubsection{Data Breach}
A data breach is a security incident at the service provider's end that can result in the disclosure of users' credentials and activity data. This was one of the main concerns for several participants, such as P\_18: \quotes{Because we heard that Google and even Apple were hacked, and thousands, even hundreds of thousands, of millions of personal information of those involved were leaked}.

%% file: sec/discussion.tex
\section{Discussion}
In this section, we summarize our results and discuss implications and recommendations stemming from our findings, summarized as follows:

\subsection{Summary of Findings}
Our findings reveal themes and use cases in which activity logs were perceived and used positively. Participants perceived some benefits in Google's \textit{Activity controls} such as, history \& bookmarking, synchronization, parental control, and security and safety. On the other hand, lack of knowledge about Google's \textit{Activity controls}, privacy concerns, biased content, and ads were on the negative side of the activity logs. Tailored content was a double-edged sword, with participants perceiving it both positively and negatively. Some of our themes for the positive aspects (history and tailored content such as personalized ads and improved suggestions) and the negative aspect (privacy concerns) align with Farke et al.'s findings~\cite{farke21}. Our analysis also revealed that participants had misconceptions, including confusion between local and online activity logs. At the extreme negative end (the disastrous), unauthorized access to activity data, data breaches at companies, and technology abuse were outstanding concerns. Such a balanced view of the activity logs and the themes we uncovered represents a valuable source for subsequent studies in this area. Moreover, our results allowed us to provide practical recommendations for policy makers and technology designers. 

In what follows, we present insights and recommendations to mitigate the risks associated with the negative aspects and maximize the benefits of users' activity logs. We close the section with a summary of the study implications and limitations.

\subsection{Implement Additional Layers of Security} \label{sec:layer}
Our participants expressed significant concerns regarding data breach and unauthorized access to their accounts. This is in line with previous work where participants had similar concerns regarding Google data collection~\cite{farke21}. We also inferred the potential risk of abusing the \textit{Activity controls} feature, for example, by a violent intimate partner, for stalking or exposing data about their victims. Clearly, accounts that save activity logs, such as Google's \textit{Activity controls}, that contain very personal details, such as \textit{Web \& App Activity}, \textit{YouTube History}, or \textit{Location History}, pose additional serious concerns. 

Thus, we recommend service providers add a mandatory additional layer of security to protect users' activity logs from serious potential risks. This includes requiring an extra verification step to access activity logs in users' accounts. For example, similar to changing a password in most service providers, requiring additional verification for the password or second-factor authentication. Moreover, we recommend service providers employ a form of feedback nudges~\cite{acquisti17}, to keep the user informed about any changes made to their ativity logs settings, such as enabling or disabling them, to reduce risks.

As of this writing, Google provides a feature for extra verification to manage (e.g., view and delete) the activity logs (see~\autoref{fig:extra_verif} in~\autorefappendix{app:extra_verif}). However, it is not enabled by default, buried inside each activity page, meaning that it requires user's prior knowledge about the \textit{Activity controls}, leaving unaware users vulnerable to risks, which may present serious concerns for them. This leads us to discuss privacy by default in the next section.

\subsection{Implement Privacy by Default} \label{sec:default} 
Research showed that default settings influence users' choices~\cite{alashwali25,baek14,bellman01}. However, service providers often adopt permissive defaults and long retention periods for users' activity logs. This was the case in Google's \textit{Activity controls} at the time of the study, where the \textit{Web \& App Activity} and the \textit{YouTube History} were enabled by default. Most of our participants kept the default settings~\cite{alashwali24}. This is also evident in related work, such as~\cite{haque23,farke21,lau18}. At the same time, most participants in our dataset found that Google's default retention periods for all three types of activity that Google logs are longer than needed, and preferred shorter retention periods~\cite{alashwali24}. This is inline with what is reported in~\cite{malkin19} regarding smart speaker users' preferences towards voice command logs. Moreover, recent work showed that participants who used privacy interfaces with preloaded default settings (presets) had less comprehension of the choices they made~\cite{alashwali25_b}. It also showed that participants preferred a privacy interface that requires them to select each setting individually over interfaces with default settings, citing the ability to customize settings as they wish and to make informed decisions about each setting as the main reasons for their preference~\cite{alashwali25_b}. Research suggests that an opt-in approach (where users need to approve using their data) is more advantageous to protecting users' information privacy than an opt-out approach (where users need to disapprove using their data)~\cite{alashwali25_b,baek14,bellman01}. 

Thus, we recommend adopting privacy by default for settings, including shorter retention periods to maintain users' activity logs. Further research is needed to identify what constitutes good defaults, and the role data protection laws and regulations can play in developing guidelines for default settings that prioritize users' privacy, especially when the data is very personal, such as the \textit{Web \& App Activity} and \textit{YouTube History}, which are, as of this writing, turned on by default in Google's \textit{Activity controls}.

\subsection{Raise Awareness about Potential Technology Abuse} \label{sec:abuse}
Participants viewed some uses of Google's \textit{Activity controls} positively, such as parental control to review the content their children viewed and location tracking by an authorized individual for security and safety. However, it should be noted that such features are prone to abuse, such as, stalking, tracking, or unauthorized exposure of log data, had they fallen into the wrong hands, for example, by a violent intimate partner. Slupska and Tanczer stressed the importance of threat modeling technology abuse in intimate partner violence, and the need for a combination of technical and social measures to address it~\cite{slupska21}. Our identification of the activity logs abuse theme is a step in this direction.  

Identifying the potential abuse of the activity logs supports our aforementioned recommendations regarding the need for a second layer of security for the activity logs feature~(\autoref{sec:layer}) to protect users' privacy by default~(\autoref{sec:default}), at least for sensitive data such as location and web activity logs, which have been reported as more sensitive than other types of logs, such as the \textit{YouTube History}~\cite{alashwali24}. Additionally, from an awareness perspective, service providers need to educate users through informational messages about the potential for abuse of these double-edged features, so that users are aware that sharing an account's password entails not only mailbox access but also access to much more personal data. Privacy experts also need to raise awareness about technology abuse, including the abuse of subtle features such as the activity logs. 

\subsection{Address Misconceptions} \label{sec:miscon}
We identified misconceptions among participants regarding intrinsic issues related to activity logs, including local versus online logs, private browsing, and VPNs. Google Chrome added information below the \quotes{Delete browsing history data} option, clarifying which data is being deleted. For example, when a Google Chrome user deletes the browsing history data while signed out, the following statement appears: \quotes{To delete browsing data from all of your synced devices and your Google Account, sign in}. However, Google account users who use browsers other than Google's Chrome, such as Mozilla's Firefox, may not be aware that deleting browsing local history does not delete online activity logs saved in the cloud, such as Google's \textit{Web \& App Activity}. 

We recommend that service providers, such as browser and VPN vendors, address these misconceptions regarding activity logs and privacy protection, for example, by providing users with more information. In addition, raising public awareness of the nuanced issues surrounding web privacy can help protect users' privacy by reducing unintentional data sharing.

\subsection{Cultural Influence} \label{sec:culture}
Our participants are from Saudi Arabia, which may be culturally, religiously, politically, and economically different than most of the WEIRD (Western, Educated, Industrialized, Rich, and Democratic) countries~\cite{hasegawa24}. This may limit the generalizability of our results. One might ask whether and how the themes we identified are linked to the participants' cultural background. While this paper provides an original perspective, when reasoning about the identified themes (summarized in~\autoref{fig:diagram}), we do not find any that may not be applicable to other cultures. At first glance, we suspected that the reported use cases of using Google's \textit{Web \& App Activity} and \textit{YouTube History} for parental control are uniquely related to the collective nature of Saudi society and the controlling parenting style that is arguably prevalent in Saudi Arabia~\cite{dwairy06}. %, as this theme is seldomly mentioned in the literature. 
However, a study in the UK identified technology repurposing (using technology for purposes beyond its original purpose) in IoT device context, e.g., using smart cameras, smart lights, and smart speakers for parental control and entertainment~\cite{chalhoub21}.

Alashwali and Cranor's observation in~\cite{alashwali24}, which produced the data we used in our analysis, whose analysis focused on participants' awareness, attitudes, preferences, concerns, and behaviors, noted that their identified themes are similar to those found in studies in the US. In this perspective, we reiterate their observation and note that we do not observe themes unique to the Saudi Arabian culture in our analysis of the positive and negative aspects and misconceptions regarding Google's \textit{Activity controls}. However, both of our studies are qualitative, thus we are limited to reasoning about the emerged themes. Further quantitative studies may reveal cultural differences at other levels, such as prevalence and scale. Moreover, it is important to consider the type of privacy control in our study, which is user-to-institution, with the institution being a company, where the cultural aspects may have less influence than in user-to-user (between a user and other users) privacy control~\cite{alashwali25}. 

\subsection{Implications} \label{sec:implications}
From a theoretical perspective, our results provide a balanced view regarding users' activity logs from users' perspective, represented by the Google \textit{Activity controls} as a case study. Namely, we distilled sub-themes that fall into the following four themes: the good, the bad, the misconception, and the disastrous (depicted in~\autoref{fig:diagram}). Our analysis highlighted overlooked themes, such as the positive perception of Google's \textit{Web \& App Activity} for parental control and account security monitoring, and the \textit{Location History} for physical safety. Additionally, we also inferred the potential abuse of the activity logs, such as in intimate partner violence.

In terms of practical implications, our results provide practical recommendations for service providers, security and privacy researchers and experts, and users alike. This includes emphasizing the need for an additional layer of security to activity logs, and for security and privacy by default, especially for data that is often perceived as sensitive, including  \textit{Web \& App Activity}. We also recommend feedback nudges that notify the account holder of changes to the activity log settings, such as when they are turned on or off. Moreover, we recommend that products, such as web browsers, include more information by design to address misconceptions about activity logs, including information about where the logs' data will be stored (locally vs. in the cloud). Finally, we recommend public awareness programs that educate users about the potential for abuse of technology features, such as activity logs, and strategies for protection and support.

\subsection{Limitations} \label{sec:limitation}
Our study has the following limitations. First, a common limitation in secondary analysis is that the data were not collected for the research question we address in this paper~\cite{doolan09}. Nevertheless, the data are closely related to our research question and have been carefully assessed for their suitability. Second, the interview data we used were collected in 2021. We balanced the cost of gathering new data versus the benefits of reusing existing data. Accordingly, we decided to reuse existing data as we do not anticipate that users' privacy perceptions change rapidly. Third, the interview data are for 30 participants living in Saudi Arabia, which may affect the generalizability of our results. However, the results of Alashwali and Cranor suggest that, Saudi participants exhibited patterns in perceptions and behaviors similar to those of US participants in similar studies with respect to user-to-institution privacy~\cite{alashwali24}. Moreover, a sample size of 30 is deemed acceptable for qualitative interview studies that focus on surfacing themes rather than drawing quantitative conclusions~\cite{distler21}. Fourth, participants' demographics are biased toward those located in the western region of Saudi Arabia (93\%) and slightly towards females (63\%). However, as noted in~\cite{alashwali24}, the western region of Saudi Arabia (a.k.a. the Makkah Region) is very large and diverse, containing several cities and towns with a population of 8,021,463 according to the latest population estimates of the Makkah region as of 2022~\cite{gas22_pop}. Fifth, participants' answers are prone to recall bias and social desirability. However, this is a common limitation in interviews and self-reporting studies. Both of these limitations were addressed in~\cite{alashwali24}. To address recall bias, the interviews employed interactive task-based questions in which participants were instructed to open their Google account's \textit{Activity controls} and answer accordingly. To address social desirability bias, participants were reminded at the beginning and during the interview that there is no right or wrong answer and that the research team has no relationship with Google. Moreover, the interviewer did not use the \quotes{security} and \quotes{privacy} words until the very last sections of the interview, to avoid privacy and security priming. Sixth, the direct quotes from participants in this paper are the most accurate translations from Arabic into English. We tried to translate participants’ responses as-is, including imperfections, accurately. Nevertheless, \quotes{lost in translation} expressions may have occurred. Finally, some of the terms used by Google during the interviews have been updated (some terms were updated during interviews with four participants~\cite{alashwali24}). However, as noted in~\cite{alashwali24}, interface and terminology changes by service providers during and after field studies are common and reported in privacy studies~\cite{malkin19,wang14,almuhimedi15}.

%% file: sec/conclusion.tex
\section{Conclusion}
In this paper, we present a secondary analysis of interview data from 30 participants in Saudi Arabia who hold Google personal accounts regarding their perceptions of Google's \textit{Activity controls}. Using template analysis, we qualitatively analyzed the data from the lens of four main themes: the good, the bad, the misconception, and the disastrous aspects of users' activity logs from the users' perspective. Unlike previous work that emphasizes the negative side of activity logs, our work provides a balanced view of users' perceptions regarding activity logs and covers the positive, negative, and extremely negative aspects, as well as the misconceptions about users' activity logs, providing a better understanding of users' perceptions of users' activity logs and a useful source for subsequent studies in related topics.

%% file: sec/ack.tex
\section{Acknowledgment}
The author acknowledges using an  AI-assisted tool (Grammarly~\cite{grammar26}) to proofread the paper to improve the English language write-up. The text is entirely generated by the author. The language corrections made by the tool were thoroughly reviewed by the author. This project was funded by the Deanship of Scientific Research (DSR) at King Abdulaziz University, Jeddah, under grant no. (GPIP: 604-612-2024). The author, therefore, acknowledges with thanks the DSR for financial support.

%% file: sec/appendices.tex
\begin{appendices}
\renewcommand{\theHfigure}{\Alph{section}\arabic{figure}}
\input{appendix/ext_results}
\clearpage
\input{appendix/questions}
\end{appendices}

%% file: appendix/ext_results.tex
\section{Additional Results}\label{app:results}
\subsection{Demographic Details}\label{app:demo}
In this section, we list the participants' demographic details obtained from~\cite{alashwali24}.
%============== DEMOGRAPHICS =====================
\begin{table}[!h]
    \small
	\centering
    \caption{Participants' general demographics. Source:~\cite{alashwali24}.}
    \label{tab:demo}
%    \resizebox{0.5\textwidth}{!}{%
	\begin{tabular}{llr}
		\toprule
		\multicolumn{3}{c}{\textit{N = 30}} \\
		\hline
		Nationality		& 	\#	&	\%		\\ 
		\hline
		\quad Saudi 	& 25 & (83\%) \\
		\quad Other    	& 5  & (17\%)  \\
		\hline
		%------------------------------------------------------------------------
		Gender			 & 	\#	&	\%	\\
		\hline
		\quad Female 	& 19 & (63\%) \\
		\quad Male    	& 11 & (37\%)  \\
		\hline
		%------------------------------------------------------------------------ 
		Residence region of Saudi Arabia			 & 	\#	&	\%	\\
		\hline
		\quad Western 	 						 & 28 & (93\%) \\
		\quad Eastern	 						 & 0 & (0\%) \\
		\quad Central    						 & 2 & (7\%) \\
		\quad Northern	 						 & 0 & (0\%) \\
		\quad Southern	 						 & 0 & (0\%) \\
		\quad Other (*required: please specify)	 & 0 & (0\%) \\
		\hline 
		%------------------------------------------------------------------------
		Age				&  \#	& \% \\
		\hline
		\quad 18 to 24		& 10 & (33\%) \\
		\quad 25 to 34		& 10 & (33\%) \\
		\quad 35 to 44		& 5 & (17\%) \\
		\quad 45 to 54 		& 5 & (17\%) \\
		\quad 55 to 64 		& 0 & (0\%) \\
		\quad 65 or above	& 0 & (0\%) \\
		\hline 
		%------------------------------------------------------------------------
		Highest degree completed &  \#	& \% \\
		\hline
		\quad Doctorate								&  2   &   (7\%)      \\
		\quad Master's								&  9   &   (30\%)      \\
		\quad Bachelor								&  11  &   (37\%)      \\
		\quad High school							&  8   &   (27\%)      \\
		\quad Intermediate school					&  0   &   (0\%)      \\
		\quad Elementary school						&  0   &   (0\%)      \\
		\quad Other (*required: please specify)		&  0   &   (0\%)      \\
		\hline
		%------------------------------------------------------------------------
		Have CS/IS/IT/CE background?				&  \#	& \% \\
		\hline
		\quad Yes							& 8 & (27\%)\\
		\quad No							& 22 & (73\%) \\
		\hline
		%------------------------------------------------------------------------
		Employment status							&  \#	& \% \\
		\hline
		\quad Student								& 11  & (37\%) \\
		\quad Full-time employee					& 11  & (37\%) \\
		\quad Part-time employee					& 0  & (0\%) \\
		\quad Self-employed or business owner		& 2  & (7\%) \\
		\quad Full-time house wife/husband			& 1  & (3\%) \\
		\quad Unemployed \& looking for a job		& 5  & (17\%) \\
		\quad Unemployed \& not looking for a job   & 0  & (0\%) \\
		\quad Unable to work						& 0  & (0\%) \\
		\quad Retired								& 0  & (0\%) \\
		\quad Other (*required: please specify)		& 0  & (0\%) \\
        \hline
%------------------------------------------------------------------------
		Google account age		& 	\#	&	\%		\\ 
		\hline
		\quad At least since a month  				& 0 & (0\%) \\
		\quad At least since 3 months					& 1 & (3\%) \\
		\quad At least since 6 months					& 2 & (7\%)  \\
		\quad At least since a year					& 2 & (7\%) \\
		\quad At least since 2 years					& 4 & (13\%) \\
		\quad At least since 3 years					& 4 & (13\%) \\
		\quad At least since 5 years					& 4 & (13\%) \\
		\quad At least since more than 5 years 		& 13 & (43\%) \\
		\quad Other (*required: please specify)		& 0  & (0\%) \\
  		\hline
  %------------------------------------------------------------------------
		Browser type		& 	\#	&	\%		\\ 
		\hline
		\quad Google Chrome 		& 22 & (73\%) \\ 
		\quad Firefox				& 2	 & (7\%) \\
		\quad Brave				& 0	 & (0\%) \\
		\quad MS Edge				& 1  &  (3\%) \\
		\quad Safari				& 5  & (17\%) \\
		\quad Opera 				& 0  & (0\%) \\
		\quad Other (*required: please specify)	& 0  & (0\%) \\
		\bottomrule
	\end{tabular}
%} % end resizebox
\end{table}
\clearpage 
%================================
\subsection{Google's \textit{Activity Controls} Extra Verification Feature}\label{app:extra_verif}
\begin{figure}[!h]
	\centering
	\includegraphics[width=0.8\textwidth]{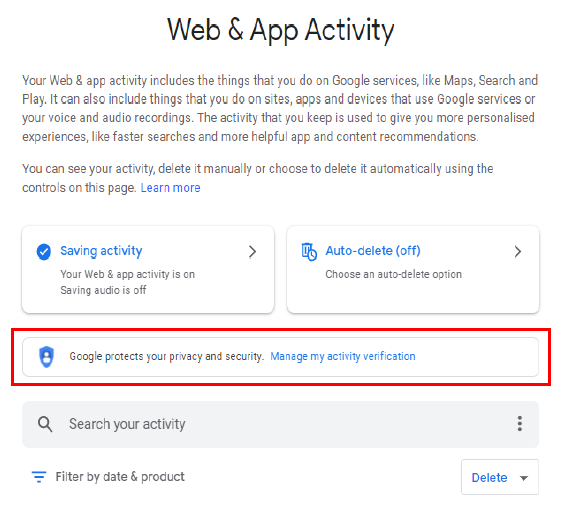}
	\caption{The extra verification feature in the \textit{Web \& App Activity}~\cite{google21}.}
	\label{fig:extra_verif}
\end{figure}

\autoref{fig:extra_verif} illustrates the extra verification feature to manage the \textit{Web \& App Activity}. The same feature is also available on the other logs: the \textit{YouTube History} and \textit{Location History}.

%=========== BEGIN ACTIVITY CONTROLS FIG =====================
%=========== END ACTIVITY CONTROLS FIG =====================

%% file: appendix/questions.tex
\section{Interview Questions} \label{app:questions}
In this section, we list the interview questions related to this paper, obtained from~\cite{alashwali24}. Interviews were conducted in Arabic. The following version is a translated version from Arabic to English. Text between square brackets was not shown to participants (added for clarification). There are more instructions and figures, and demographic, debriefing, and other minor questions, which we omit here for brevity~\cite{alashwali24}.%The full interview script is available upon request.
%================================================
\subsection{Users' Experiment}
\begin{figure}[!h]
	\centering
	\includegraphics[width=0.8\columnwidth]{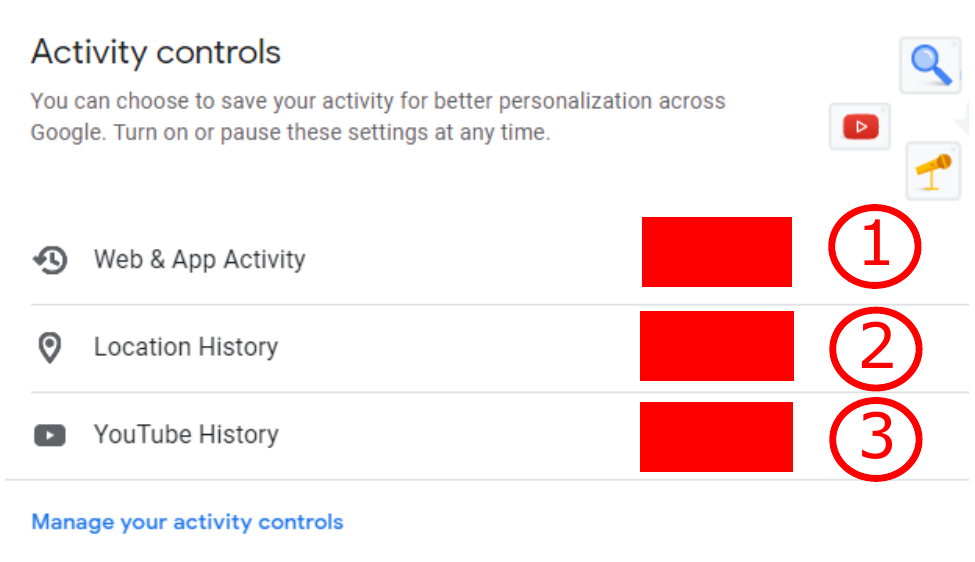}
	\caption{The Activity Controls basic settings~\cite{google21}. Participants were asked to report their accounts' current settings (either \quotes{on} or \quotes{paused}). We covered our settings in the figure to avoid bias or confusion. Source:~\cite{alashwali24}.}
	\label{fig:activity_controls}
\end{figure}

%\hfill\newline 
[Questions about awareness of Google's Activity Controls]
\begin{itemize}
% Q. 1
    \item \textbf{Q. 1:} Which choice appears to you in the shaded areas numbers: 1, 2, and 3, in front of the following items: Web \& App Activity, Location History, and YouTube History? \par
    \textbf{Answers:} \begin{inparaitem}[$\circ$]\item \quotes{Stop,}\footnote{\label{note1}A fictitious answer was added as an answer validity check.} \item \quotes{On,} \item \quotes{Paused}\end{inparaitem} \par 
    \textbf{Related figures:}  \autoref{fig:activity_controls} 
    \vspace{4pt}
%------------------------------------------------
% Q. 2
    \item \textbf{Q. 2:} Were you aware of Google's Activity Controls that enable you to control the activities that Google saves about you such as the Web \& App Activity, YouTube History, and Location History? (yes/no, elaborate if possible)? \par 	
    \textbf{Answers:} 
    \begin{inparaitem} [$\circ$] \item \quotes{Yes,} \item \quotes{No,} \item \quotes{I heard about such a thing,} \item \quotes{I expected there is such a thing} \end{inparaitem} \vspace{4pt}
%-------------------------------------------------	
% Q. 3
	\item \textbf{Q. 3:} \textbf{[If Q. 2 answer is \quotes{yes,} \quotes{I heard about such a thing,} or \quotes{I expected there is such a thing}]} How did you know about, heard of, or expected the existence of Google's Activity Controls? \par 
	\textbf{Answers:} Open	
\end{itemize}
%================================================
[Questions about usage of Google's Activity Controls]
%\subsubsection{Usage of Google's Activity Controls}
\begin{itemize}
% Q. 4
	\item \textbf{Q. 4:} Have you ever used Google's Activity Controls? (yes/no, elaborate if possible). \par 
	\textbf{Answers:} \begin{inparaitem} [$\circ$] \item \quotes{Yes,} \item \quotes{No,} \item \quotes{I do not remember}\end{inparaitem}
 \vspace{4pt}
%-------------------------------------------------	
% Q. 5
	\item \textbf{Q. 5:} \textbf{[If Q. 4 answer is \quotes{yes}]} Why did you use Google's Activity Controls? \par 
	\textbf{Answers:} Open  
 \vspace{4pt}
%-------------------------------------------------	
% Q. 6
	\item \textbf{Q. 6:} \textbf{[If Q. 4 answer is \quotes{yes}]} What changes did you make using Google's Activity Controls? \par 
	\textbf{Answers:} Open 
 \vspace{4pt}
%-------------------------------------------------	
% Q. 7
	\item \textbf{Q. 7:} \textbf{[If Q. 4 answer is \quotes{no}]} If you have not used the Activity Controls before, it means you are keeping the default settings regarding the activities that Google saves about you, such as the Web \& App Activity and the YouTube History. What are your reasons for not using Google's Activity Controls and keeping the default settings? \par 
	\textbf{Answers:} Open 
\end{itemize}	

\begin{figure}[!h]
	\centering
	\includegraphics[width=0.8\columnwidth]{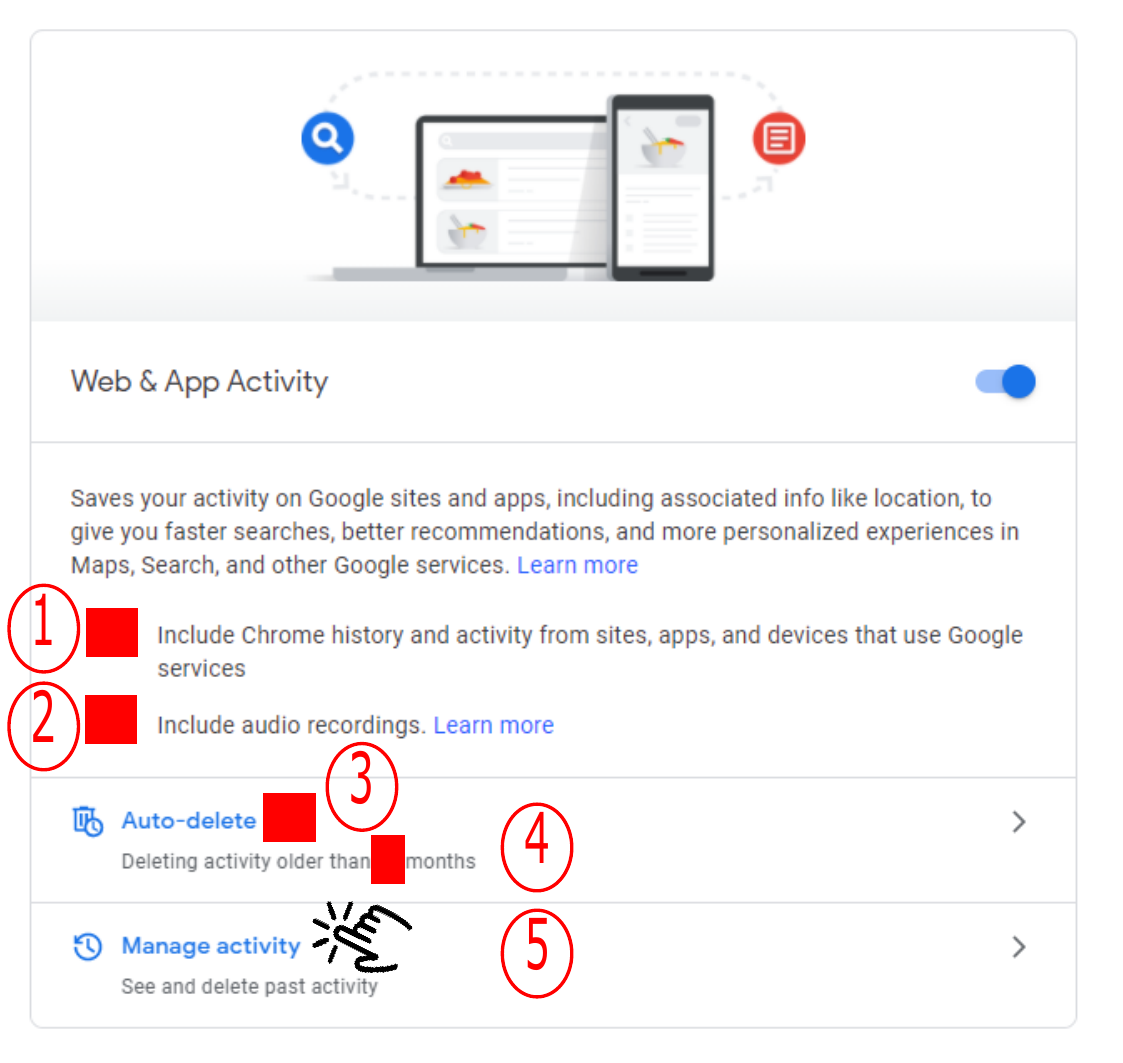}
	\caption{The advanced options of the Web \& App Activity~\cite{google21}. Participants were asked to report their settings (either checked \quotes{\cmark} or unchecked). We covered our settings in the figure to avoid bias or confusion. Source:~\cite{alashwali24}.}
	\label{fig:web_advanced}
\end{figure}

\begin{figure} [!h]
	\centering\includegraphics[width=0.8\columnwidth]{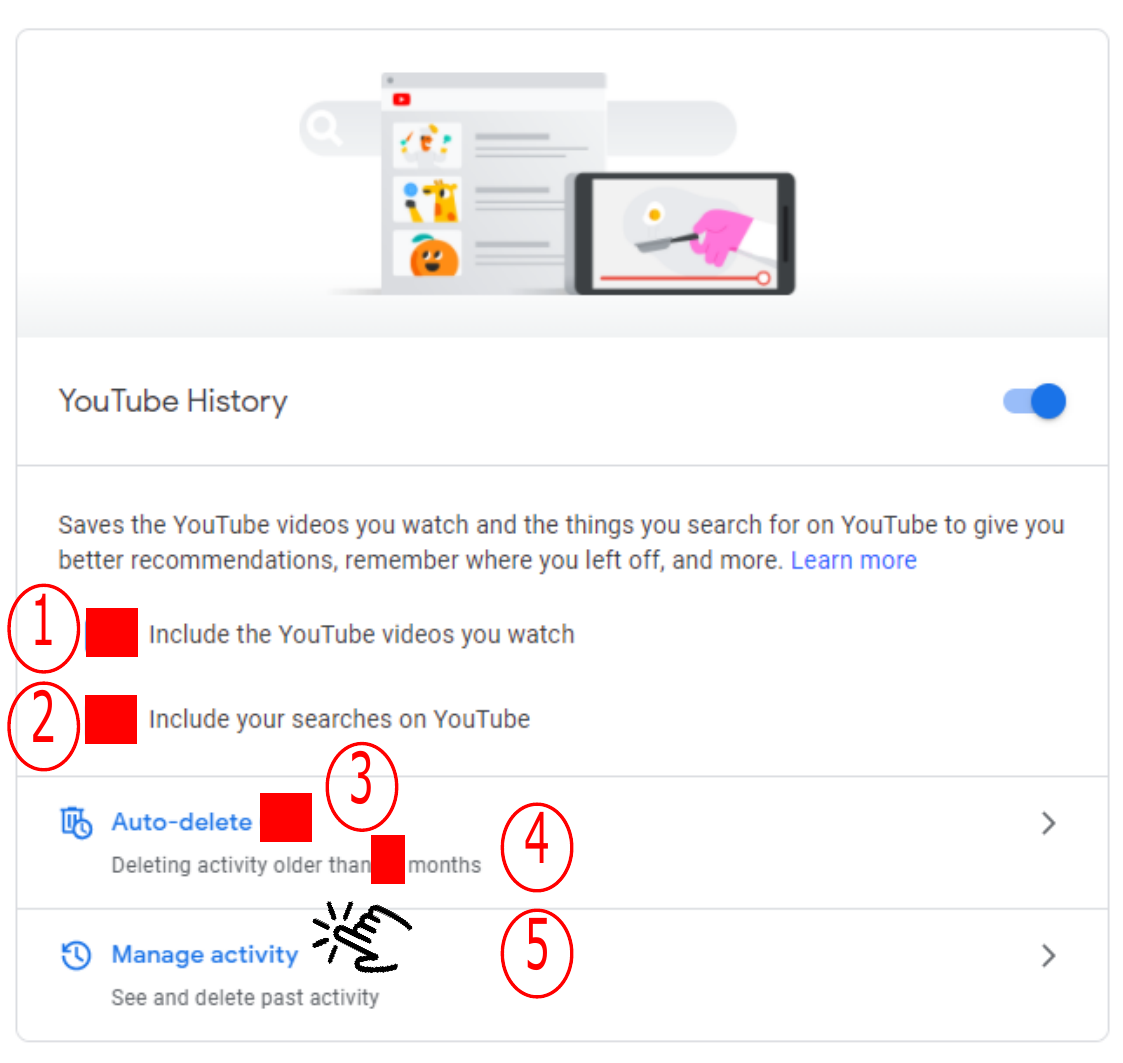}
	\caption{The advanced options of the YouTube History~\cite{google21}. Participants were asked to report their settings (either checked \quotes{\cmark} or unchecked). We covered our settings in the figure to avoid bias or confusion. Source:~\cite{alashwali24}.}
	\label{fig:youtube_advanced}
\end{figure}

%================================================
%\subsubsection{Preferences and Attitudes Towards Google's Data Practices and Activity Controls}
[Questions about preferences and attitudes towards Google's data practices and Activity Controls]
\begin{itemize}
% Q. 8
    \item \textbf{Q. 8:} \textbf{[If Q. 1 answer is at least one \quotes{on}]} What are the checked choices in your [Web \& App Activity $|$ YouTube History $|$ Location History] \footnote{\label{note2}The participant was asked about either the Web \& App Activity, YouTube History, or Location history, depending on the participants' answer on the basic settings (Q. 1). Full details of the task description can be found in~\cite{alashwali24}}? \par
    \textbf{Answers:} \par
    \textbf{[If we proceeded to the Web \& App Activity advanced settings, the following answers were shown]} \par 
    \begin{inparaitem}[$\square$] \item \quotes{Include Chrome history and activity from sites, apps, and devices that use Google services,} \item \quotes{Include audio recordings,} \item \quotes{None of the above,} \item  \quotes{Other (*required: please specify)}\end{inparaitem}  
     \vspace{2pt}
     
    \textbf{[If we proceeded to the YouTube History advanced settings, the following answers were shown]} \par 
    \begin{inparaitem}[$\square$] \item \quotes{Include the YouTube videos you watch,} \item \quotes{Include your searches on YouTube,} \item \quotes{None of the above,} \item  \quotes{Other (*required: please specify)}\end{inparaitem} \par 
    \textbf{Related figures:} \autoref{fig:web_advanced} and~\autoref{fig:youtube_advanced} 
     \vspace{2pt}
     
    \textbf{[If we proceeded to the Location History advanced settings]}\par 
    [No choices were presented as the Location History has no advanced options] 
    \vspace{4pt}
    %------------------------------------------------
% Q. 9    
    \item \textbf{Q. 9:} \textbf{[If Q.1 answer is at least one \quotes{on}]} What is the status of the Auto-delete that appears to you\footref{note2}? \par
    \textbf{Answers:} \begin{inparaitem} [$\circ$] \item \quotes{On,} \item \quotes{Off,} \item \quotes{Not applicable,} \item \quotes{Stop,}\footref{note1} \item \quotes{Other (*required: please specify)} \end{inparaitem}\par 
    \textbf{Related figures:} \autoref{fig:web_advanced} and~\autoref{fig:youtube_advanced} 
    \vspace{4pt}
    %------------------------------------------------
    
    \item \textbf{Q. 10:} \textbf{[If Q. 9 answer is \quotes{on}]} How long does Google save your data before automatic deletion?
    \footref{note2}? \par
    \textbf{Answers:} \begin{inparaitem} [$\circ$] \item \quotes{3 months,} \item \quotes{18 months,} \item \quotes{36 months,} \item \quotes{Other (*required: please specify)} \end{inparaitem} \par 
    \textbf{Related figures:} \autoref{fig:web_advanced} 
    \vspace{4pt}
    %------------------------------------------------
    
	\item \textbf{Q. 11:} \textbf{[If Q.1 answer is at least one \quotes{on}]} Have you found [previous searches or so $|$ geographic locations] from the [Web \& App Activity $|$ YouTube History $|$ Location History]\footref{note2}? (yes/no, elaborate if possible). \par 
	\textbf{Answer:} \begin{inparaitem} [$\circ$] \item \quotes{Yes,} \item {no} \end{inparaitem}\par 
	\textbf{related Figures:} \autoref{fig:web_advanced} and~\autoref{fig:youtube_advanced}
 \vspace{4pt}

%-------------------------------------------------		
	\item \textbf{Q. 12:} \textbf{[If Q.11 answer is \quotes{yes}]} Describe your feeling after you found [Web \& App Activity $|$ YouTube History $|$ Location History]\footref{note2} data about you? \par 
	\textbf{Answers:} Open
 \vspace{4pt}
	
%-------------------------------------------------	
\item \textbf{Q. 13:} \textbf{[If Q.11 answer is \quotes{yes}]} Were you aware that Google saves the [Web \& App Activity $|$ YouTube History $|$ Location History]\footref{note2} data about you? \par

\textbf{Answer:} \begin{inparaitem}[$\circ$]  \item \quotes{Yes,} \item \quotes{No,} \item \quotes{I expected that, but I am not certain,} \item \quotes{I heard about that, but I am not certain}\end{inparaitem}
\vspace{4pt}

%-------------------------------------------------
\item \textbf{Q. 14:}  \textbf{[If Q. 13 answer is \quotes{no}]} Do you have a suggestion if Google implemented, it would have informed you about the saving of your data? \par 
\textbf{Answers:} Open
\vspace{4pt}

%-------------------------------------------------
\item \textbf{Q. 15:}  \textbf{[If Q. 13 answer is \quotes{yes}]} Did you know that you can review the [Web \& App Activity $|$ YouTube History $|$ Location History]\footref{note2} data that Google saves about you? \par
\textbf{Answers:} \begin{inparaitem}[$\circ$] \item \quotes{Yes,} \item \quotes{No}\end{inparaitem} 
\vspace{4pt}

%-------------------------------------------------
\item \textbf{Q. 16:} \textbf{[If Q.15 answer is \quotes{yes}]} Approximately, how often do you review these data? \par 
\textbf{Answers:} \begin{inparaitem}[$\circ$] \item \quotes{At least once a day,} \item \quotes{At least once a week,} \quotes{At least once a month,} \quotes{At least once every 3 months,} \quotes{At least once every 6 months,} \quotes{At least once a year,} \quotes{Less than once a year,} \quotes{I never reviewed them,} \quotes{I do not remember}\end{inparaitem}
\vspace{4pt}

%-------------------------------------------------
\item \textbf{Q. 17:} \textbf{[If Q.13 answer is \quotes{yes}]} Did you know that you can manually delete some or all of the [Web \& App Activity $|$ YouTube History $|$ Location History]\footref{note2} data that Google saves about you? 
\vspace{4pt}

%-------------------------------------------------
\item \textbf{Q. 18:} \textbf{[If Q.17 answer is \quotes{yes}]} Approximately, how often do you manually delete these data? \par 
\textbf{Answers:} \begin{inparaitem}[$\circ$] \item \quotes{At least one a day,} \item \quotes{At least once a week,} \item \quotes{At least once a month,} \item \quotes{At least once every 3 months,} \item \quotes{At least once every 6 months,} \item \quotes{At least once a year,} \item \quotes{Less than once a year,} \item \quotes{I never deleted them,} \item \quotes{I do not remember}\end{inparaitem} \vspace{4pt}
%-------------------------------------------------

\item \textbf{Q. 19:} \textbf{[If Q.13 answer is \quotes{yes}]} Did you know about the Auto-delete feature which allows you to specify how long Google saves the [Web \& App Activity $|$ YouTube History $|$ Location History]\footref{note1} data before they are automatically deleted? (yes/no, elaborate if possible) \par 
\textbf{Answers:} \begin{inparaitem}[$\circ$] \item \quotes{Yes,} \item \quotes{No}\end{inparaitem}
\end{itemize}
%-------------------------------------------------	
\subsection{Auto-delete}
\begin{itemize}
% Q. 21
	\item \textbf{Q. 20:} Do you think that the default retention period specified by Google (18 months, i.e. 1.5 years) to save the Web \& App Activity data at Google before auto-deletion is suitable? (yes/no, elaborate if possible) \par 
	\textbf{Answers:} \begin{inparaitem}[$\circ$] \item \quotes{Yes,} \item \quotes{No}\end{inparaitem} 
\vspace{4pt}
%-------------------------------------------------
% Q. 22		
	\item\textbf{Q. 21} \textbf{[If Q. 20 answer is \quotes{no}]} What period do you suggest for saving the Web \& App Activity data before Google automatically deletes them? \par 
	\textbf{Answers:} Open 
 \vspace{4pt}
%-------------------------------------------------
% Q. 23		
	\item \textbf{Q. 22:} Do you think that the default retention period specified by Google (36 months, i.e. 3 years) to save the YouTube History data is suitable? (yes/no, elaborate if possible) \par 
	\textbf{Answers:} \begin{inparaitem}[$\circ$] \item \quotes{Yes,} \item \quotes{No}\end{inparaitem} 
 \vspace{4pt}
%-------------------------------------------------	
% Q. 24	
	\item\textbf{Q. 23} \textbf{[If Q. 22 answer is \quotes{no}]} What period do you suggest for saving the YouTube History data at Google before Google automatically deletes them? \par 
	\textbf{Answers:} Open  
 \vspace{4pt}
%-------------------------------------------------	
% Q. 25	
	\item \textbf{Q. 24:} Do you think that the default retention period specified by Google (18 months, i.e. 1.5 years) to save the Location History data is suitable? (yes/no, elaborate if possible) \par
 	\textbf{Answers:} \begin{inparaitem}[$\circ$] \item \quotes{Yes,} \item \quotes{No}\end{inparaitem} 
  \vspace{4pt}
%-------------------------------------------------
% Q. 26		
	\item\textbf{Q. 25} \textbf{[If Q. 24 answer is \quotes{no}]} What period do you suggest for saving the Location History data at Google before Google automatically deletes them? \par 
	\textbf{Answers:} Open 
 \vspace{4pt}
 \end{itemize}
%-------------------------------------------------
\subsection{Your Data Privacy and Sensitivity}
\begin{itemize}
% Q. 27		
	\item \textbf{Q. 26:} If Google saves the following data: Web \& App activity, Location History, and YouTube History. Rank them according to the degree of privacy (no. 1 High Privacy, 2 Medium, 3 Low). Note: you can choose more than one item with an equal degree of privacy. \par 	
	\textbf{Answers:} \begin{inparaitem}[$\circ$] \item \quotes{1 (High),} \item \quotes{2 (Medium),} \item \quotes{3 (Low)} \end{inparaitem} 
 \vspace{4pt}
%-------------------------------------------------	
% Q. 28
	\item \textbf{Q. 27:} To what extent do you consider your [Web \& App Activity, YouTube History, Location History]\footnote{Asked as 3 separate questions, one question per data type.} data at Google services sensitive? \par 
	\textbf{Answers:} \begin{inparaitem}[$\circ$] \item \quotes{They are all sensitive,} \item \quotes{Some are sensitive,} \item \quotes{All are insensitive,} \item \quotes{Other (*required: please specify)} \end{inparaitem} 
 \vspace{4pt}
 \end{itemize}
%-------------------------------------------------	
\subsection{Acceptability for Saving Your Web \& App Activity}
\begin{itemize}
% Q. 29
	\item \textbf{Q. 28:} To what extent would you accept if Google saves, processes, and analyzes your Web \& App activity data, for the purpose of improving the services provided to you, such as speeding up your searches, providing you with better recommendations, and more personalized experiences at Google services? \par 
	\textbf{Answers:} \begin{inparaitem}[$\circ$] \item \quotes{Very acceptable,} \item \quotes{Somewhat acceptable,} \item \quotes{Neutral,} \item \quotes{Somewhat unacceptable,} \item \quotes{Completely unacceptable} \end{inparaitem} 
 \vspace{4pt}
%-------------------------------------------------	
% Q. 30
	\item \textbf{Q. 29:} to what extent would you accept if Google saves, processes, and analyzes your Web \& App activity data, for the purpose of displaying ads to you on Google services such as Google search engine or YouTube? \par 
	\textbf{Answers:} \begin{inparaitem}[$\circ$] \item \quotes{Very acceptable,} \item \quotes{Somewhat acceptable,} \item \quotes{Neutral,} \quotes{Somewhat unacceptable,} \item \quotes{Completely unacceptable}  \end{inparaitem} 
 \vspace{4pt}
%-------------------------------------------------	
% Q. 31
	\item \textbf{Q. 30:} To what extent would you accept if Google saves, processes, and analyzes your Web \& App activity data, for the purpose of displaying ads to you on non-Google websites and apps that have a partnership with Google (such as newspaper websites and gaming apps)? \par 
	\textbf{Answers:} \begin{inparaitem}[$\circ$] \item \quotes{Very acceptable,} \item \quotes{Somewhat acceptable,} \item \quotes{Neutral,} \item \quotes{Somewhat unacceptable,} \item \quotes{Completely unacceptable} \end{inparaitem} 
\end{itemize}	

%================================================
\subsection{Privacy Concerns}
\begin{itemize}
	% Q. 32
	\item \textbf{Q. 31:} Previously, did you have any concerns regarding your privacy while using Google services, such as the Google search engine, Google Maps, or YouTube? \par
	\textbf{Answers:} \begin{inparaitem}[$\circ$] \item \quotes{Yes,} \item \quotes{No,} \item \quotes{Neutral} \end{inparaitem} 
 \vspace{4pt}
%-------------------------------------------------		
% Q. 33
	\item \textbf{Q. 32:} \textbf{[If Q. 31 answer is \quotes{yes}]} If your answer is yes, what are these concerns? \par 
	\textbf{Answers:} Open 
 \vspace{4pt}
%-------------------------------------------------	
% Q. 34
	\item \textbf{Q. 33:} Have you ever taken any steps to protect your privacy online while using Google services, such as the Google search engine, Google Maps, or YouTube? \par 
	\textbf{Answers:} Open with \begin{inparaitem}[$\circ$] \item \quotes{Yes,} \item \quotes{No}\end{inparaitem}
 \vspace{4pt}
%-------------------------------------------------	
% Q. 35	
	\item \textbf{Q. 34:} [If Q. 33 answer is \quotes{yes}] If your answer is yes, what are these steps? \par 
	\textbf{Answers:} Open
 \vspace{4pt}
%-------------------------------------------------	
% Q. 36		
	\item \textbf{Q. 35:} Do you plan to take any additional steps to protect your privacy online while using Google services, such as the Google search engine, Google Maps, or YouTube? \par
	\textbf{Answers:} Open with \begin{inparaitem}[$\circ$] \item \quotes{Yes,} \item \quotes{No}\end{inparaitem}
 \vspace{4pt}
%-------------------------------------------------
% Q. 37	
	\item \textbf{Q. 36:} [If Q. 35 answer is \quotes{yes}] If your answer is yes, what are these steps? \par 
	\textbf{Answers:} Open
\end{itemize}

%-------------------------------------------------
%=========== BEGIN ACTIVITY CONTROLS FIG =====================
%=========== END ACTIVITY CONTROLS FIG =====================

%=========== BEGIN ADVANCED WEB SETTINGS FIG =====================
%=========== END ADVANCED WEB SETTINGS FIG =====================

%=========== BEGIN ADVANCED YOUTUBE SETTINGS FIG =====================
%=========== END ADVANCED YOUTUBE SETTINGS FIG =====================